\documentclass[pss]{wiley2sp} 
\usepackage{amsmath}

\begin{document}

\title{Competition and/or Coexistence of Antiferromagnetism and Superconductivity in CeRhIn$_5$ and CeCoIn$_5$}

\titlerunning{CeRhIn$_5$ and CeCoIn$_5$}

\author{%
  Georg Knebel\textsuperscript{\Ast,1},
  Dai Aoki\textsuperscript{1},
  Jean-Pascal Brison\textsuperscript{1},
  Ludovic Howald\textsuperscript{1},
  Gerard Lapertot\textsuperscript{1},
  Justin Panarin\textsuperscript{1},
  Stephane Raymond\textsuperscript{1},
  Jacques Flouquet\textsuperscript{1}}

\authorrunning{Georg Knebel et al.}

\mail{e-mail
  \textsf{georg.knebel@cea.fr}, Phone:
  +33-4-38783951, Fax: +33-4-38785096}

\institute{Commissariat \`a l' \'Energie Atomique, INAC, SPSMS, 17 rue des Martyrs, 38054 Grenoble, France
  \textsuperscript{1}\,}

\received{XXXX, revised XXXX, accepted XXXX} 
\published{XXXX} 


\abstract{%
%
%
%
\abstcol{%
  The Ce compounds CeCoIn$_5$ and CeRhIn$_5$ are ideal model systems to study the competition of antiferromagnetism (AF) and superconductivity (SC). Here we discuss the pressure--temperature and magnetic field phase diagrams of both compounds. In CeRhIn$_5$ the interesting observation is that in zero magnetic field a coexistence AF+SC phase exist inside the AF phase below the critical pressure $p_{\rm c}^\star \approx 2$~GPa. Above $p_{\rm c}^\star$ AF is suppressed in zero field but can be re-induced by applying a magnetic field. The collapse of AF under pressure coincides with the abrupt change of the Fermi surface.  
  }{%
In CeCoIn$_5$ a new phase appears at low temperatures and high magnetic field (LTHF) which vanishes at the upper critical field $H_{\rm c2}$. In both compounds the paramagnetic pair breaking effect dominates at low temperature. We discuss the evolution of the upper critical field under high pressure of both compounds and propose a simple picture of the glue of reentrant magnetism to the upper critical field in order to explain the interplay of antiferromagnetic order and superconductivity. 
}
}

%
%

\maketitle   

\section{Pressure--temperature phase diagrams}

\begin{figure}[t]%
\includegraphics*[width=0.9\linewidth]{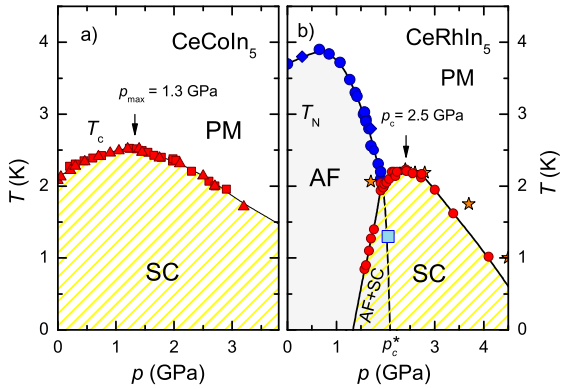}
\caption{%
a) Pressure temperature phase diagram of CeCoIn$_5$ at zero magnetic field from specific heat and ac susceptibility measurements \cite{Knebel2004}. b) Pressure temperature phase diagram of CeRhIn$_5$ at zero magnetic field from specific heat measurements ~\cite{Knebel2006}. The square indicates  $T_{\rm N}$ for $p=2.05$~GPa $>p_{\rm c}^\star$ observed in recent NQR experiments \cite{Yashima2007} 
}
\label{PD}
\end{figure}

The competition and/or coexistence of long range antiferromagnetism (AF) and unconventional superconductivity (SC) has been studied intensively in Ce heavy fermion compounds, notably in the Ce 115 family. This family offers the possibility to study the interplay of the different ground states by tuning the system either by pressure ($p$), or by magnetic field ($H$).
CeCoIn$_5$ is superconducting at ambient pressure and no magnetism appears at zero magnetic field and on applying pressure. The observed non Fermi liquid properties in the normal state coupled to the strong coupling of  SC  indicate the closeness to a quantum critical point in CeCoIn$_5$. Under pressure, the superconducting transition temperature $T_{\rm c}$ has a maximum around $p_{\rm max} = 1.3$~GPa and the superconducting domain extends up to at least $p = 5$~GPa \cite{Sidorov2002,Knebel2004} (see Fig.~\ref{PD}a). The general picture for the high pressure phase diagram of CeCoIn$_5$ is that with increasing pressure the critical antiferromagnetic fluctuations are suppressed and an usual Fermi-liquid ground state is observed above $p \approx 1.5$~GPa. The increase of $T_{\rm c}$ may be explained by the increase of the heavy fermion band width \cite{Yashima2004}.

CeRhIn$_5$ orders  at ambient pressure below $T_{\rm N} = 3.8$~K in an incommensurate antiferromagnetic structure with an ordering vector $q_{\rm ic}=$(1/2, 1/2, 0.297). Under pressure, $T_{\rm N} (p)$ has a smooth maximum around 0.7~GPa and in zero magnetic field AF is rapidly suppressed above $p_{\rm c}^\star \approx 2$~GPa (see Fig.~\ref{PD}b)). At the pressure $p_{\rm c}^\star$ the N\'eel temperature $T_{\rm N}$ and the superconducting transition temperature $T_{\rm c}$ coincides. For $p>p_{\rm c}^\star$ the opening of the superconducting gap $\Delta_{\rm SC}$ prevent the formation of long range magnetic order in zero field. Below $p_{\rm c}^\star$ coexistence of AF and SC is reported \cite{Park2006,Knebel2006,Yashima2007}. However, the transition to the AF+SC state appears inhomogeneous and the anomaly of the superconducting transition in the specific heat is very tiny.  In neutron scattering experiments it is observed that the magnetic structure of CeRhIn$_5$ is modified under pressure and the magnetic ordering vector at $p = 1.7$~GPa has changed to $q_{\rm ic}=$(1/2,~1/2,~0.4) \cite{Raymond2008,Aso2009}. The ordered magnetic moment can be estimated to be lower than 0.2 $\mu_{\rm B}$. The ground state AF+SC at lowest $T$ was proved to be homogeneous by NQR indicating the microscopic coexistence of AF and SC \cite{Yashima2007,Yashima2009}. The remarkable observation is that the magnetic ordering is changing from incommensurate to commensurate with $q_{\rm c}=$ (1/2,~1/2,~1/2) at $p\sim 1.7$~GPa in the uniformly coexisting phase AF+SC.
Fortunately NQR experiments can be performed in better hydrostatic conditions than neutron scattering measurements and thus is seems reasonable to believe that below $p_{\rm c}^\star$ the transition from AF to AF+SC  is associated with a simultaneous magnetic transition from incommensurate to commensurate AF and a SC transition. This leads to a  specific heat anomaly  very far from the usual BCS behavior. 
 Tiny traces of the commensurate phase have been observed in the NQR spectra even at lower pressures and this can explain the observation of superconducting transitions in resistivity or ac susceptibility which has been also reported \cite{Chen2006,Paglione2008}. Thus AF and SC coexist only if AF has a commensurate structure. A mismatch between incommensurate magnetic ordering and the lattice periodicity is not compatible with an homogeneous AF+SC state. 
Above $p_{\rm c}^\star$ the ground state is purely superconducting with $d$ wave symmetry \cite{Mito2001}. A drastic change of the Fermi surface at the critical pressure $p_{\rm c}$ indicates the transition from an 4$f$ localized to a $4f$ itinerant state \cite{Shishido2005}. This change coincides with the maximum of the residual resistivity and the effective mass \cite{Knebel2008}. The shape of the paramagnetic Fermi surface at high pressure is very close to that of CeCoIn$_5$ at ambient pressure. 

\section{Magnetic phase diagrams}

\begin{figure}[t]%
\includegraphics*[width=\linewidth]{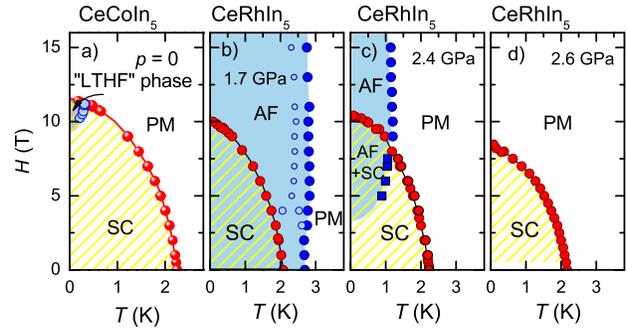}
\caption{%
a) Magnetic field--temperature phase diagram of CeCoIn$_5$ at zero pressure obtained by ac calorimetry for $H\parallel a$. The microscopic origin of the ''LTHF'' phase is still under debate; surely, it has a magnetic and superconducting component. b-d)
$H-T$ phase diagram of CeRhIn$_5$ at $p=1.7$~GPa ($p<p_{\rm c}^\star$), 2.4~GPa ($p_{\rm c}^\star<p<p_{\rm c}$), and 2.6~GPa ($p>p_{\rm c}$) for magnetic fields in the $ab$-plane.} 
\label{field_dependence}
\end{figure}

The $H-T$ phase diagram of CeCoIn$_5$ is plotted in Fig.~\ref{field_dependence}a) for a magnetic field $H \parallel a$. It shows the peculiarity that  the superconducting transition gets first order below $T_0 \approx 0.7$~K $\approx 0.3 T_{\rm c}$ \cite{Tayama2002,Bianchi2002,Bianchi2003b,Radovan2003,Kakuyanagi2005,Tayama2005,Mitrovic2006} for both field directions. This first order transition is either due to strong paramagnetic effects or due to a strong polarization of the Ce atoms in high magnetic fields \cite{Buzdin2007}. Furthermore at low temperature and high magnetic field a new phase (LTHF) appears inside the superconducting state which was first claimed to be a Fulde-Ferrel Larkin-Ovchinnikov (FFLO) state (an overview is given in Ref.~\cite{Matsuda2007}). SC in CeCoIn$_5$ seems favorable for the formation of such a state. However, recent neutron scattering experiments clearly show that the LTHF phase  has at least a magnetic component as incommensurate magnetic ordering has been observed \cite{Kenzelmann2008}. Astonishingly, the upper field limit of the LTHF phase coincides with the SC upper critical field $H_{\rm c2}$ and the magnetic Bragg peaks disappear abruptly at $H_{\rm c2}$. Under  pressure the area of the LTHF state expands and appears at higher temperatures and lower magnetic fields and the first order transition is observed at higher temperatures \cite{Miclea2006}.

 At low field, SC appears in CeCoIn$_5$ out of a non-Fermi liquid regime. Only for $H > H_{\rm c2}$ Fermi-liquid behavior is observed and a magnetic field induced quantum critical point may be achieved at $H_{\rm QCP} \approx H_{\rm c2}$ for both field directions \cite{Bianchi2003c,Ronning2005}. Detailed very low temperature transport measurements clearly show that the Fermi-liquid temperature does not vanish exactly at $H_{\rm c2}$, but inside the SC state \cite{Howald2009} as has been suggested from Hall effect measurements \cite{Singh2007}. Under pressure $H_{\rm QCP}$ vanishes around $p_{\rm max}\approx 1.3$~GPa \cite{Ronning2006} and a clear separation between $H_{\rm QCP}$ and $H_{\rm c2}$ has been observed. If $H_{\rm QCP}$ is connected to some hidden magnetic ordered state this would be suppressed at $p_{\rm max}$. 

The $(H, T)$ phase diagram of CeRhIn$_5$ is shown in Fig.\ref{field_dependence}b-d) for different pressures. Below $p_{\rm c}^\star$  SC appears inside the AF ordered state. Here we have plotted the superconducting phase detected by resistivity measurements which may not probe bulk SC. The magnetic phase diagram seems to be unchanged in comparison to $p=0$ and SC appears on top of the AF phases. For $p>p_{\rm c} \approx 2.5$~GPa magnetic order can not be observed anymore. The interesting case appears for $p_{\rm c}^\star<p<p_{\rm c}$. In zero field the ground state is purely superconducting, but under the application of an external magnetic field, not  only SC is suppressed, but also AF order reenters inside the SC state \cite{Park2006,Knebel2006}. The idea is that AF nucleates inside the vortex cores and long range order would be promoted thanks to superconductivity. This re-entrance phase collapses at $p_{\rm c}$. The microscopic nature of the coexistence AF+SC state is not determined up to now, neither for the AF+SC state at zero field below $p_{\rm c}^\star$ nor for the field induced state in the pressure range $p_{\rm c}^\star<p<p_{\rm c}$. No anomaly can be detected when the $T_{\rm N} (H)$ and the $H_{\rm c2}(T)$ lines are crossing. 
One remarkable difference is that in CeRhIn$_5$ the magnetic order is preserved far above $H_{\rm c2}$ but in CeCoIn$_5$ the field induced state is limited to the upper critical field. 

\section{Upper critical field and superconducting phase diagram}

\begin{figure}[t]%
\includegraphics*[width=0.9\linewidth]{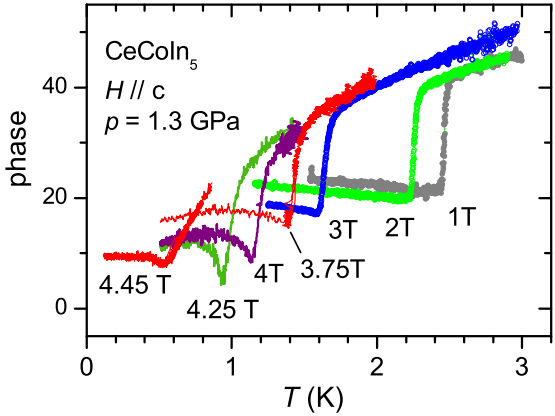}
\caption{%
Typical form of the obtained signal in the phase of the ac signal obtained for CeCoIn$_5$ at $p=1.3$~GPa for a magnetic field $H\parallel c$-axis. The transition gets very peaked below $T_0 \approx 1.2$~K, which may indicates the change to a first order transition.} 
\label{phase13}
\end{figure}

More detailed information on the superconducting state can be obtained from an analysis of the upper critical field $H_{\rm c2}$. Generally, $H_{\rm c2}$  is determined by orbital and paramagnetic pair breaking effects. The orbital effect can be estimated from the initial slope 
$H'_{c2}=({\rm d}H_{\rm c2}/{\rm d}T)$ of $H_{\rm c2}$ at $T_{\rm c}$ using $H_{\rm orb} = -0.7 H'_{c2}T_{\rm c}$ \cite{Werthammer1966}. As close to $T_{\rm c}$ the orbital effect is always dominating over the paramagnetic effect, which would have an infinite slope at $T_{\rm c}$, the measured initial slope can also be used to estimate the averaged effective mass of the carriers perpendicular to the applied magnetic field as $-H'_{c2} \propto (m^\star) ^2 T_{\rm c}$. 

The paramagnetic pair breaking effect originates from Zeeman splitting of the single electron levels and its value can be estimated by $H_{\rm P} = \sqrt{2} \Delta_{\rm SC} / g\mu_B$ with $\Delta_{\rm SC}$ being the value of the superconducting gap at $T=0$ and $g$ the gyromagnetic ratio. The main difficulty in evaluating $H_{\rm P}$ concerns the correct estimation of the $g$ factor which can  significantly deviate from the value $g=2$ of a free electron due to spin-orbit coupling or the exchange with local moments. Furthermore, for superconductors where strong coupling corrections are important (which is the case for the Ce 115 family), the enhancement of the coupling parameter $\lambda$ leads to a weaker paramagnetic limitation. As discussed previously \cite{Knebel2008} for CeCoIn$_5$ and CeRhIn$_5$ strong coupling has to be taken into account to give a correct fit of $H_{\rm c2}$ over the whole temperature range up to $T_{\rm c}$. 

In the following we discuss qualitatively the pressure evolution of $H_{\rm c2}$.

\begin{figure}[t]%
\includegraphics*[width=\linewidth]{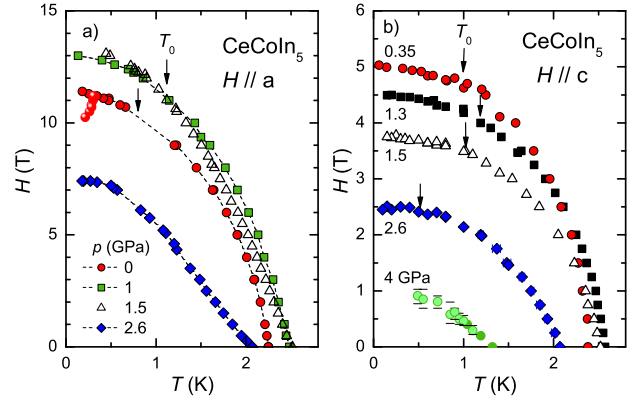}
\caption{%
Upper critical field of CeCoIn$_5$ for magnetic field $H\parallel a$ (a) and $H\parallel c$. Arrows indicate $T_0$ where the transition changes to  first order at high magnetic fields. A nice signature of the LTHF phase has only been observed for $H\parallel a$ at zero pressure.
} 
\label{CeCoIn5_Hc2}
\end{figure}

\subsection{CeCoIn$_5$}

$H_{\rm c2}$ of CeCoIn$_5$ has been measured by ac  calorimetry in a diamond anvil high pressure cell with argon as pressure medium. Measurements for different field directions are performed in the same pressure cell which can be rotated by 90$^\circ$ inside a superconducting magnet with 13.5~T maximal field. Typical experimental data are shown in Fig.~\ref{phase13}. From the sharp transition in the phase of the measured thermocouple voltage the $H_{\rm c2}$ phase diagram can be plotted. Figure~\ref{CeCoIn5_Hc2}  shows $H_{\rm c2} (T)$ for different pressures up to 4~GPa for both field directions. At low pressure we find good agreement with previously published data \cite{Tayama2005,Miclea2006}. In difference to \cite{Miclea2006}, we do not find a clear signature of the LTHT phase in our experiment. Thus we will not to discuss its variation under pressure in detail. However, we can detect the first order transition at high magnetic fields for $H\parallel a$ ($H_{\rm c2}^a$) up to 1.5~GPa and for $H\parallel c$ ($H_{\rm c2}^c$) up to 2.6~GPa. In CeCoIn$_5$ the anisotropy of the critical field $H_{\rm c2}^a / H_{\rm c2}^c$ for $T\to 0$ increases from  2.2 at $p=0$ to 3 at 2.6~GPa.

\begin{figure}[t]%
\includegraphics*[width=\linewidth]{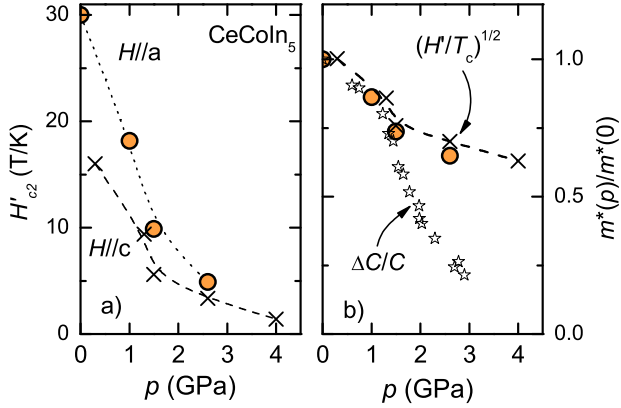}
\caption{%
a) Pressure dependence of the initial slope of CeCoIn$_5$ for field $H \parallel a$ (circles) and $H \parallel c$ (crosses). b) Normalized effective mass determined from the initial slope $m^\star \propto \sqrt{H'/T_{\rm c}}$ (weak coupling approximation) for $H \parallel a$ (circles) and $H \parallel c$ (crosses) as function of pressure. Furthermore the jump of the specific heat $\Delta C/C$ at $T_{\rm c}$ normalized to its value at $p=0$ from \cite{Knebel2004} is plotted (stars).
} 
\label{initial_slope}
\end{figure}

The pressure dependence of the initial slope $H'$ is shown for different field directions in Fig.~\ref{initial_slope}a). 
$H'$ decreases monotonously with pressure for both field directions. No anomaly appears close to $p_{\rm max}$. This shows that CeCoIn$_5$ is pushed away from any quantum critical point under pressure and antiferromagnetic fluctuations are continuously suppressed \cite{Yashima2004}. From the initial slope we can estimate that the effective mass which is plotted normalized to its value at zero pressure in Fig.~\ref{initial_slope}b). It decreases almost by a factor of 2 for both field directions. This is consistent with the drop of the cyclotron masses of the $\alpha$ branches and the $\beta$ branch of the cylindrical Fermi surfaces which decrease from $15m_0$ to $7m_0$ and from $60m_0$ to $40m_0$ at 3~GPa, respectively \cite{Shishido2003}. In Fig.~\ref{initial_slope}b) we compare the $p$ dependence of $H'/T_{\rm c}$  with that of the jump of the specific heat at $T_{\rm c}$ from ref.~\cite{Knebel2004}. $\Delta C/C =5$ at T$_{\rm c}$ for $p=0$ and decreases to 1 for $p=3$~GPa which is the value for a BCS superconductor with $d$ wave gap. This large jump $\Delta C/C$ at $p=0$ can be explained by a strong coupling of the SC order parameter to a fluctuating magnetization originated by a suppressed putative AF transition below $T_{\rm c}$ \cite{Kos2003}. Up to $p\approx 1.5$~GPa the relative $p$ dependence of $\sqrt{H'/T_{\rm c}}$ and  $\Delta C/C$ at $T_{\rm c}$ normalized to its zero pressure value scale very well, but for higher pressures, the normalized $\Delta C/C$ decreases much faster. Remarkably, this corresponds to the pressure where $H_{\rm QCP}\to 0$. This may indicate that the AF fluctuations are strongly suppressed for $p>p_{\rm max}$ and (the coupling between the fluctuations and the SC order parameter becomes almost inefficient at high pressures and $\Delta C/C$ is no more determined by the strong coupling, but reaches the constant BCS value of weak coupling limit.

\begin{figure}[t]%
\includegraphics*[width=\linewidth]{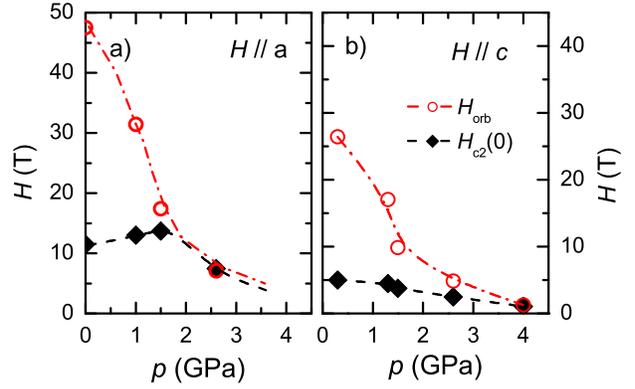}
\caption{%
Pressure dependence of the orbital limit $H_{\rm orb} = -0.7 H' T_{\rm c}$ and upper critical field $H_{\rm c2}(0)$ for $T\to 0$ for $H\parallel a$ and  $H\parallel c$. Lines are guides to the eyes.
} 
\label{limits}
\end{figure}

In Fig.~\ref{limits} we compare at $T=0$ the theoretical orbital critical field $H_{\rm orb}$ with the experimental data extrapolated to 0~K as function of pressure. For $H\parallel c$ the measured critical field is monotonously decreasing with increasing pressure. At $p=4$~GPa we extrapolate $H_{\rm c_2} (0) \approx 1$~T. Contrary, $H_{\rm c2}(0)$ increases as $T_{\rm c}$ for $H\parallel a$ and goes through a maximum at 1.3~GPa. 
The correct determination of the Pauli limiting field from a fit of $H_{\rm c2}$ with theoretical models is difficult. Previously we found $g$ = 4.7 at $p=0$ for $H\parallel c$  and $g$ = 7 for $p=1.34$~GPa whereas for $H\parallel a$ a $g$-factor around 2 has been observed \cite{Knebel2008}. There is no simple explanation of the strong enhancement of $g$ for $H\parallel c$. However, it is obvious that at low pressure $H_{\rm c2}(0)$ is dominated by the paramagnetic effect for both field directions, and for $H\parallel c$ this effect is even much more pronounced. 
For $H\parallel c$ the orbital critical field decreases monotonously as the initial slope and is of the order of the experimental value for $p\approx  2.5$~GPa.  For $H\parallel a$ the orbital critical field $H_{\rm orb}$ is at $p\sim 1.5$~GPa on the same order than $H_{\rm c2}(0)$ due to the strong suppression of the initial slope under pressure. From these data it gets obvious that the Maki parameter $\alpha = \sqrt{2}H_{\rm orb}/H_{\rm c2}(0)$ is strongly pressure dependent and decreases for $H\parallel a$ from 4.4 to 1.34 at 2.6~GPa, for $H\parallel c$ the drop from 7.4 to 1.8 is even stronger, thus under high pressure the appearance of a sole superconducting FFLO state is unlikely.

\subsection{Comparison with CeRhIn$_5$}

\begin{figure}[t]%
\includegraphics*[width=0.9\linewidth]{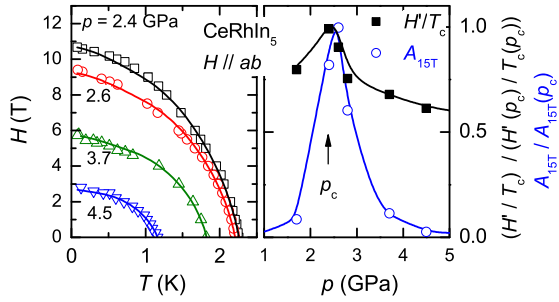}
\caption{
a) Upper critical field of CeRhIn$_5$ for field in the $ab$ plane for different pressures. b) Pressure variation of $H'/T_{\rm c}$ (squares) and the $A$ coefficient (circles) of the resistivity measured at 15~T normalized to their values at the critical pressure $p_{\rm c}$ for CeRhIn$_5$. 
} 
\label{CeRhIn5_Hc2}
\end{figure}

Finally, we compare the $p$ dependence of $H_{\rm c2}$ of CeCoIn$_5$ to that of CeRhIn$_5$. Fig.~\ref{CeRhIn5_Hc2}a) shows $H_{\rm c2}$ of CeRhIn$_5$ for $H\parallel ab$ plane obtained from previous resistivity measurements \cite{Knebel2008}.  The values of $T_{\rm c}$ and $H_{\rm c2}$ are comparable to CeCoIn$_5$ at ambient pressure.
From the initial slope the pressure dependence of the effective mass can be estimated as shown in Fig.\ref{CeRhIn5_Hc2}b). In difference to CeCoIn$_5$ the maximum of $H'$ coincides with the maximum of $T_{\rm}$. However, the pressure dependence of the $A$ coefficient of the resistivity measured under a magnetic field of 15~T has a much stronger pressure dependence. This strong $p$ dependence of $A$ above $p_{\rm c}$ may be  due to a possible change of the degeneracy of Ce associated with the crossing of a valence quantum critical point \cite{Watanabe2009} under pressure. This will affect strongly the transport properties, e.g. the Kadowaki Woods ratio will change by a factor of 15 entering in the intermediate valence regime \cite{Knebel2008}. 

It is worthwhile to remark that despite a quite similar topology of their paramagnetic Fermi surfaces, the CeRhIn$_5$ and CeCoIn$_5$ differ drastically in the anisotropy of $H_{\rm c2}$. $H'_{\rm c2}$ is less anisotropic in CeRhIn$_5$ but $H_{\rm c2}^a < H_{\rm c2}^c $ in difference to CeCoIn$_5$  \cite{Ida2008}.  This strongly indicates that the effective mass build by the correlations is very sensitive to the replacement of Rh by Co and the interplay between the 4$f$ electrons of Ce and 3$d$ or 4$d$ electrons of the transition metal which differ strongly on their spatial extensions. 

Concerning the LTHF phase of CeCoIn$_5$ in comparison to the field induced AF+SC in CeRhIn$_5$ a simple picture which emerges is the interplay between the superconducting gap $\Delta_{\rm SC}$ and a pseudogap $\Delta_{\rm AF}$ which is built through the establishment of the itinerant AF order. Magnetic field will modify the SC gap and also the evolution of the spin resonance observed in neutron scattering \cite{Panarin2009}. Close to $H_{\rm c2}$ $\Delta_{\rm SC}$ may have a form like the pseudogap shape  due to the appearance of vortices favorable for the recovery of AF, which may be hidden by superconductivity.  In absence of superconductivity CeCoIn$_5$ at $p=0$ is assumed to be an AF with $T_{\rm N} < T_{\rm c}$ and a critical magnetic field $H_{\rm M} < H_{\rm c2} (0)$. The strength of $\Delta_{\rm SC}$ prevent the establishment of AF in zero field at $p=0$. Under magnetic field $\Delta_{\rm SC}$ decreases, but at some field $H_{\rm LTHF}$ 
it will be of the same order as the pseudogap $\Delta_{\rm AF}$, and AF will reenter and be stabilized by the SC gap up to $H_{\rm c2} (0)$. Above $H_{\rm c2}$ AF in CeCoIn$_5$ is not possible as AF needs the SC gap.  A test of the possible validity of this simple image would be to follow the interplay of $\Delta_{\rm SC}$ and $\Delta_{\rm AF}$ under pressure. If $\Delta_{\rm SC} (p)$ decreases under pressure as the strong coupling may get less important (at least above 1.3~GPa $T_{\rm c}(p)$ decreases) the condition $\Delta_{\rm SC} (H), =\Delta_{\rm AF} (H)$ under pressure may be realized for a smaller value of $H/H_{\rm c2}$ than at $p=0$, in good agreement with the previous experimental observation \cite{Miclea2006}. 
Obviously, the AF correlations collapses around $p_{\rm max}$ and thus it is expected that the LTHF phase disappears at this pressure. 
 
The complex $(H, T, p)$ phase diagram of both systems is governed by the coupling of critical fluctuations to the superconducting order parameter. 
In zero field, both systems, CeCoIn$_5$ already at $p=0$ and CeRhIn$_5$ above $p_c^\star$, correspond to the case where AF order is suppressed due to SC. However, a magnetic field  leads to a renaissance of AF. 


\begin{acknowledgement}
This project has been supported by the ANR projects ECCE and Nemsicom. 
\end{acknowledgement}

%

\begin{thebibliography}{[10]}

\bibitem{Sidorov2002}
 \textsc{V.\,A. Sidorov},  \textsc{M.~Nicklas},  \textsc{P.\,G. Pagliuso},
  \textsc{J.\,L. Sarrao},  \textsc{Y.~Bang},  \textsc{A.\,V. Balatzky},  and
  \textsc{J.\,D. Thompson},
 \jr{Phys. Rev. Lett.} \textbf{89}, 157004 (2002).


\bibitem{Knebel2004}
 \textsc{G.~Knebel},  \textsc{M.\,A. Measson},  \textsc{B.~Salce},
  \textsc{D.~Aoki},  \textsc{D.~Braithwaite},  \textsc{J.\,P. Brison},  and
  \textsc{J.~Flouquet},
 \jr{J. Phys.: Condens. Matter} \textbf{16}, 8905 (2004).


\bibitem{Yashima2004}
 \textsc{M.~Yashima},  \textsc{S.~Kawasaki},  \textsc{Y.~Kawasaki},
  \textsc{G.~q~Zheng},  \textsc{Y.~Kitaoka},  \textsc{H.~Shishido},
  \textsc{R.~Settai},  \textsc{Y.~Haga},  and  \textsc{Y.~\=Onuki},
 \jr{J. Phys. Soc. Jpn.} \textbf{73}, 2073 (2004).


\bibitem{Knebel2006}
 \textsc{G.~Knebel},  \textsc{D.~Aoki},  \textsc{D.~Braithwaite},
  \textsc{B.~Salce},  and  \textsc{J.~Flouquet},
 \jr{Phys. Rev. B} \textbf{74}, 020501(R) (2006).


\bibitem{Yashima2007}
 \textsc{M.~Yashima},  \textsc{S.~Kawasaki},  \textsc{H.~Mukuda},
  \textsc{Y.~Kitaoka},  \textsc{H.~Shishido},  \textsc{R.~Settai},  and
  \textsc{Y.~\=Onuki},
 \jr{Phys. Rev. B} \textbf{76}, 020509 (2007).


\bibitem{Park2006}
 \textsc{T.~Park},  \textsc{F.~Ronning},  \textsc{H.\,Q. Yuan},  \textsc{M.\,B.
  Salamon},  \textsc{R.~Movshovich},  \textsc{J.\,L. Sarrao},  and
  \textsc{J.\,D. Thompson},
 \jr{Nature} \textbf{440}, 65 (2006).


\bibitem{Raymond2008}
 \textsc{S.~Raymond},  \textsc{G.~Knebel},  \textsc{D.~Aoki},  and
  \textsc{J.~Flouquet},
 \jr{Phys. Rev. B} \textbf{77}, 172502 (2008).


\bibitem{Aso2009}
 \textsc{N.~Aso},  \textsc{K.~Ishii},  \textsc{H.~Yoshizawa},
  \textsc{T.~Fujiwara},  \textsc{Y.~Uwatoko},  \textsc{G.\,F. Chen},
  \textsc{N.\,K. Sato},  and  \textsc{K.~Miyake},
 \jr{Journal of the Physical Society of Japan} \textbf{78}, 073703 (2009).


\bibitem{Yashima2009}
 \textsc{M.~Yashima},  \textsc{H.~Mukuda},  \textsc{Y.~Kitaoka},
  \textsc{H.~Shishido},  \textsc{R.~Settai},  and  \textsc{Y.~Onuki},
 \jr{arXiv: 0906.0078} (2009).


\bibitem{Chen2006}
 \textsc{G.\,F. Chen},  \textsc{K.~Matsubayashi},  \textsc{S.~Ban},
  \textsc{K.~Deguchi},  and  \textsc{N.\,K. Sato},
 \jr{Phys. Rev. Lett.} \textbf{97}, 017005 (2006).


\bibitem{Paglione2008}
 \textsc{J.~Paglione},  \textsc{P.\,C. Ho},  \textsc{M.\,B. Maple},
  \textsc{M.\,A. Tanatar},  \textsc{L.~Taillefer},  \textsc{Y.~Lee},  and
  \textsc{C.~Petrovic},
 \jr{Phys. Rev. B} \textbf{77}, 100505 (2008).


\bibitem{Mito2001}
 \textsc{T.~Mito},  \textsc{S.~Kawasaki},  \textsc{G.~q~Zheng},
  \textsc{Y.~Kawasaki},  \textsc{K.~Ishida},  \textsc{Y.~Kitaoka},
  \textsc{D.~Aoki},  \textsc{Y.~Haga},  and  \textsc{\=Onuki},
 \jr{Phys. Rev. B} \textbf{63}, 220507(R) (2001).


\bibitem{Shishido2005}
 \textsc{H.~Shishido},  \textsc{R.~Settai},  \textsc{H.~Harima},  and
  \textsc{Y.~\=Onuki},
 \jr{J. Phys. Soc. Jpn.} \textbf{74}, 1103 (2005).


\bibitem{Knebel2008}
 \textsc{G.~Knebel},  \textsc{D.~Aoki},  \textsc{J.\,P. Brison},  and
  \textsc{J.~Flouquet},
 \jr{Journal of the Physical Society of Japan} \textbf{77}, 114704 (2008).


\bibitem{Tayama2002}
 \textsc{T.~Tayama},  \textsc{A.~Harita},  \textsc{T.~Sakakibara},
  \textsc{Y.~Haga},  \textsc{H.~Shishido},  and  \textsc{R.\,S.\,Y.
  \=Onuki},
 \jr{Phys. Rev. B} \textbf{65}, 180504(R) (2002).


\bibitem{Bianchi2002}
 \textsc{A.~Bianchi},  \textsc{R.~Movshovich},  \textsc{N.~Oeschler},
  \textsc{P.~Gegenwart},  \textsc{F.~Steglich},  \textsc{J.\,D. Thompson},
  \textsc{P.\,G. Pagliuso},  and  \textsc{J.\,L. Sarrao},
 \jr{Phys. Rev. Lett.} \textbf{89}, 137002 (2002).


\bibitem{Bianchi2003b}
 \textsc{A.~Bianchi},  \textsc{R.~Movshovich},  \textsc{C.~Capan},
  \textsc{P.\,G. Pagliuso},  and  \textsc{J.\,L. Sarrao},
 \jr{Phys. Rev. Lett.} \textbf{91}, 187004 (2003).


\bibitem{Radovan2003}
 \textsc{H.\,A. Radovan},  \textsc{N.\,A. Fortune},  \textsc{T.\,P. Murphy},
  \textsc{S.\,T. Hannahs},  \textsc{E.\,C.\,P.\,S.\,W. Trozer},  and
  \textsc{D.~Hall},
 \jr{Nature (London)} \textbf{425}, 51 (2003).


\bibitem{Kakuyanagi2005}
 \textsc{K.~Kakuyanagi},  \textsc{M.~Saitoh},  \textsc{K.~Kumagai},
  \textsc{S.~Takashima},  \textsc{M.\,N. andH Takaki},  and
  \textsc{Y.~Matsuda},
 \jr{Phys. Rev. Lett.} \textbf{94}, 047602 (2005).


\bibitem{Tayama2005}
 \textsc{T.~Tayama},  \textsc{Y.~Namai},  \textsc{T.~Sakakibara},
  \textsc{M.~Hedo},  \textsc{Y.~Uwatoko},  \textsc{H.~Shishido},
  \textsc{R.~Settai},  and  \textsc{Y.~\=Onuki},
 \jr{J. Phys. Soc. Jpn.} \textbf{74}, 1115 (2005).


\bibitem{Mitrovic2006}
 \textsc{V.\,F. Mitrovi\'{c}},  \textsc{M.~Horvati\'{c}},
  \textsc{C.~Berthier},  \textsc{G.~Knebel},  \textsc{G.~Lapertot},  and
  \textsc{J.~Flouquet},
 \jr{Phys. Rev. Lett.} \textbf{97}, 117002 (2006).


\bibitem{Buzdin2007}
 \textsc{A.~Buzdin},  \textsc{Y.~Matsuda},  and  \textsc{T.~Shibauchi},
 \jr{EPL} \textbf{80}, 67004 (2007).


\bibitem{Matsuda2007}
 \textsc{Y.~Matsuda} and  \textsc{H.~Shimahara},
 \jr{J. Phys. Soc. Jpn.} \textbf{76}, 051005 (2007).


\bibitem{Kenzelmann2008}
 \textsc{M.~Kenzelmann},  \textsc{T.~Strassle},  \textsc{C.~Niedermayer},
  \textsc{M.~Sigrist},  \textsc{B.~Padmanabhan},  \textsc{M.~Zolliker},
  \textsc{A.\,D. Bianchi},  \textsc{R.~Movshovich},  \textsc{E.\,D. Bauer},
  \textsc{J.\,L. Sarrao},  and  \textsc{J.\,D. Thompson},
 \jr{Science} \textbf{321}, 1652--1654 (2008).


\bibitem{Miclea2006}
 \textsc{C.\,F. Miclea},  \textsc{M.~Nicklas},  \textsc{D.~Parker},
  \textsc{K.~Maki},  \textsc{J.\,L. Sarrao},  \textsc{J.\,D. Thompson},
  \textsc{G.~Sparn},  and  \textsc{F.~Steglich},
 \jr{Phys. Rev. Lett.} \textbf{97}, 039901 (2006).


\bibitem{Bianchi2003c}
 \textsc{A.~Bianchi},  \textsc{R.~Movshovich},  \textsc{I.~Vekhter},
  \textsc{P.\,G. Pagliuso},  and  \textsc{J.\,L. Sarrao},
 \jr{Phys. Rev. Lett.} \textbf{91}, 257001 (2003).


\bibitem{Ronning2005}
 \textsc{F.~Ronning},  \textsc{C.~Capan},  \textsc{A.~Bianchi},
  \textsc{R.~Movshovich},  \textsc{A.~Lacerda},  \textsc{M.~Hundley},
  \textsc{J.\,D. Thompson},  \textsc{P.\,G. Pagliuso},  and  \textsc{J.\,L.
  Sarrao},
 \jr{Phys. Rev. B} \textbf{71}, 104528 (2005).


\othercit
\bibitem{Howald2009}
 \textsc{L.\,Howald et~al.},
to be published.


\bibitem{Singh2007}
 \textsc{S.~Singh},  \textsc{C.~Capan},  \textsc{M.~Nicklas},
  \textsc{M.~Rams},  \textsc{A.~Gladun},  \textsc{H.~Lee},  \textsc{J.\,F.
  DiTusa},  \textsc{Z.~Fisk},  \textsc{F.~Steglich},  and
  \textsc{S.~Wirth},
 \jr{Physical Review Letters} \textbf{98}(5), 057001 (2007).


\bibitem{Ronning2006}
 \textsc{F.~Ronning},  \textsc{C.~Capan},  \textsc{E.~Bauer},  \textsc{J.\,D.
  Thompson},  \textsc{J.\,L. Sarrao},  and  \textsc{R.~Movshovich},
 \jr{Phys. Rev. B} \textbf{73}, 064519 (2006).


\bibitem{Werthammer1966}
 \textsc{N.\,R. Werthammer},  \textsc{E.~Hefland},  and  \textsc{P.\,C.
  Hohenberg},
 \jr{Phys. Rev.} \textbf{147}, 295 (1966).


\bibitem{Shishido2003}
 \textsc{H.~Shishido},  \textsc{T.~Ueda},  \textsc{S.~Hashimoto},
  \textsc{T.~Kubol},  \textsc{R.~Settai},  \textsc{H.~Harima},  and
  \textsc{Y.~\=Onuki},
 \jr{J. Phys.: Condens. Matter} \textbf{15}(32), L499--L504 (2003).


\bibitem{Kos2003}
 \textsc{i.\,c.\,v. Kos},  \textsc{I.~Martin},  and  \textsc{C.\,M.
  Varma},
 \jr{Phys. Rev. B} \textbf{68}(5), 052507 (2003).


\bibitem{Watanabe2009}
 \textsc{S.~Watanabe},  \textsc{A.~Tsuruta},  \textsc{K.~Miyake},  and
  \textsc{J.~Flouquet},
 \jr{arXiv: 0906.4136} (2009).


\bibitem{Ida2008}
 \textsc{Y.~Ida},  \textsc{R.~Settai},  \textsc{Y.~Ota},  \textsc{F.~Honda},
  and  \textsc{Y.~\=Onuki},
 \jr{J. Phy. Soc. Jpn.} \textbf{77}, 084708 (2008).


\othercit
\bibitem{Panarin2009}
 \textsc{J.\,Panarin}, \textsc{S. Raymond}, \textsc{G. Lapertot}, and \textsc{J.~Flouquet},
 \jr{arXiv: 0908.2688} (2009).



\end{thebibliography}
%

\providecommand{\WileyBibTextsc}{}
\let\textsc\WileyBibTextsc
\providecommand{\othercit}{}
\providecommand{\jr}[1]{#1}
\providecommand{\etal}{~et~al.}

\end{document}